\documentclass[12pt,letterpaper,twopage,pdf]{article}
\pdfoutput=1
\usepackage[top=1.135in,bottom=1.17in,left=1.16in,right=1.16in]{geometry}

\usepackage{epsfig,amssymb}

\newcommand{\nc}{\newcommand}
\nc{\beq}{\begin{equation}}
\nc{\eeq}{\end{equation}}
\nc{\tr}{\mathrm{Tr}}
\nc{\gev}{\,\mathrm{GeV}}
\newcommand{\met}{E\!\!\!/_T}

\begin{document}

\title{TASI Lectures on Jet Substructure}

\author{Jessie Shelton\thanks
                 {jshelton@physics.harvard.edu}\\[4mm]
Yale University Physics Department and \\
Harvard University Physics Department\\
High Energy Theory Group \\
17 Oxford Street\\
Cambridge, MA 02138, USA }
\date{June, 2012}
\maketitle
\begin{abstract}

  Jet physics is a rich and rapidly evolving field, with many
  applications to physics in and beyond the Standard Model. These
  notes, based on lectures delivered at the June 2012 Theoretical
  Advanced Study Institute, provide an introduction to jets at the
  Large Hadron Collider.  Topics covered include sequential jet
  algorithms, jet shapes, jet grooming, and boosted Higgs and top
  tagging.

\end{abstract}

\clearpage

\tableofcontents

\section{Introduction}

These notes are writeups of three lectures delivered at the
Theoretical Advanced Study Institute in Boulder, Colorado, in June
2012.  The aim of the lectures is to provide students who have little
or no experience with jets with the basic concepts and tools needed to
engage with the rapidly developing ideas concerning the use of jets in
new physics searches at the LHC.  A certain amount of familiarity with
the structure of QCD, and in particular with QCD showers, is assumed.

Lecture one introduces sequential jet algorithms, and develops several
main tools in substructure analyses using the boosted Higgs as an
example.  Lecture two delves further into jet grooming and jet shapes,
and in lecture three we conclude with an overview of top tagging and
BSM searches.

\section{Lecture I: Jets, Subjets, and Sequential Jet Algorithms}

To understand jet substructure and its applications, we must first
begin by understanding jets.  Jets, together with parton distribution
functions and factorization theorems, are the phenomenological tool
that allow us to separate out the perturbatively describable hard
interactions in proton-proton collisions, and thereby enable us to
make quantitative predictions for events involving strongly
interacting particles.  Jet cross-sections necessarily depend on the
algorithm used to define a jet. There are many jet algorithms, each
one with its own strengths and weaknesses.

The first jet algorithm was developed for $e ^ + e^-\to $ hadron
events by Sterman and Weinberg in 1977 \cite{Sterman:1977wj}.  In this
algorithm events are declared to have two jets if all but a fraction
$\epsilon $ of the total energy in the event can be contained within
two cones of half-angle $\delta$. That is, radiation off of one of the
initial partons must be sufficiently hard,
\beq
\label {eq:eps}
E_{rad} >\epsilon
\eeq
and at sufficiently wide angles from either of the other jets,
\beq
\label {eq:de}
\theta_{min} > \delta
\eeq
for the radiation to be resolved as a separate jet.  How many events
have two jets and how many contain three or more obviously depends on
the exact values chosen for $\epsilon$ and $\delta$.  For all
sufficiently large $\epsilon/E_{tot}$ and $\delta$, the partonic
cross-section for radiation of an extra parton into the region of
phase space defined by Eqs.~\ref{eq:eps} and \ref{eq:de} is
sufficiently isolated from the soft and collinear singular regions of
phase space that rates and distributions can be calculated reliably in
perturbation theory. Of course, this is a partonic calculation, and to
fully match the partonic picture onto reconstructed sprays of hadrons
requires some additional theoretical machinery to describe such
effects as (for example) hadronization.  For our purposes, however, a
parton shower picture will suffice.

The Sterman-Weinberg algorithm is the ur-example of a {\it cone}
algorithm. While cone algorithms present a very intuitive picture of
parton radiation, they can be somewhat clumsy in practice,
particularly as the number of jets increases, and they are not in
active use in most experiments today.  Other algorithms can deal much
more flexibly with high jet multiplicity.  One such flexible algorithm
is the JADE algorithm, developed by the JADE collaboration in the late
1980s, also for $e^+ e^-\to$ hadrons \cite{Bartel:1986ua,
  Bethke:1988zc}.  Here, jets are constructed by iteratively
recombining final state particles.  Define a metric to measure the
separation between final state particles $i$ and $j$,
\beq
\label {eq:yJADE}
y_{ij}\equiv \frac{m_{ij}^ 2}{Q^2} \approx\frac{2 E_i E_j(1-\cos^2\theta_{ij})}{Q^2},
\eeq
where $Q $ is the total energy of the event.  Note that $y_{ij}$
vanishes if either $i$ or $j$ is soft ($E_i\to 0 $ or $E_j\to 0 $), or
if $i$ and $j$ are collinear ($\cos\theta_{ij}\to 1$).  We can now
construct jets using the following recipe:
\begin {itemize}

\item Compute the interparticle distances $y_{ij}$ for all particles
  in the final state, and find the pair $\{i,j\}$ with the minimum
  $y_{ij}$.

\item If this minimum $y_{ij}< y_0$ for some fixed parameter $y_0$,
  combine $i$ and $j $ into a new particle, and go back to the
  previous step.

\item If $y_{ij}> y_0$, declare all remaining particles to be jets.

\end {itemize}
Since clustering of particles proceeds from smaller values of $y_{ij}$
to larger values, this recipe preferentially clusters particles that
are probing the regions of phase space dominated by the soft and
collinear singularities. In a sense, the algorithm is trying to
combine hadrons into partons by making its best guess for the
reconstructed parton shower.  The JADE algorithm has only one
parameter, the separation cutoff $y_0$, and clearly can handle
different jet multiplicities in an efficient way by varying $y_0$.  It
is the ur-example of a {\it sequential recombination algorithm}, and
the ancestor of all jet algorithms in wide use at the LHC.

The most direct descendent of the JADE algorithm is the {\it $k_T$
  algorithm} \cite{Catani:1991hj}, which replaces the particle energy
factor $E_iE_j$ in the Jade metric, Eq.~\ref{eq:yJADE}, with the
factor $\min (E_i^2, E_j^2)$:
\beq y_{ij} = \frac{2 \min (E_i^2, E_j^2) (1-\cos^2\theta_{ij})}{Q^2}.
%            \underrightarrow{\theta_{ij}\to 0} \frac{ k_{\perp}^2}{Q^2} .
\label{eq:kTee}
\eeq
This still ensures that the metric goes to zero when either $E_i \to
0$ or $E_j \to 0$ are soft, but has the advantage that the relative
softness of a particle depends only on its own energy, and not that of
the other particle in the pair.  This fixes up a technical drawback to
the JADE algorithm, where $y_{ij}^{(JADE)}\propto E_i E_j$ allows two
very soft particles to be combined even if they are at very wide
angles from each other.  Using $y_{ij}\propto \min (E_i^2, E_j^2)$
means soft particles will get preferentially clustered with
nearby harder particles instead.

For small $\theta_{ij}$, the numerator of Eq.~\ref{eq:kTee} can be
written as simply $ k_{\perp}^2$, the transverse momentum of the
softer particle relative to the harder particle---hence the name of
the algorithm.  In this form the metric is directly related to QCD
splitting functions.

To create a version of the $k_T$ algorithm that can be used at hadron
colliders, where the total energy $Q^2$ is unknown, both the algorithm
and the metric have to be adapted \cite{Ellis:1993tq}.  In the metric,
we simply use longitudinally boost-invariant quantities $p_T$ and
$\Delta R$ instead of $E$ and $\cos\theta_{ij}$, and let the metric
become dimensionful,
\beq 
d_{ij} = \frac{ \min (p_{T,i}^2, p_{T,j}^2) \Delta R_{ij}^2}{R^2} .
\eeq
The angular parameter $R$ introduced here will replace $y_0$ as
determining the cutoff for combining particles, as we will see.  We
need in addition to define the quantities
\beq 
d_{iB} = p_{T,i}^2
\eeq
for each particle $i$, since  we need to also consider splittings from the
beam.

The recombination algorithm now works as follows:
\begin {itemize}

\item Compute $d_{ij} $ and $d_{iB}$ for all particles in the final
  state, and find the minimum value.

\item If the minimum is a $d_{iB}$, declare particle $i$ a jet, remove
  it from the list, and go back to step one.

\item If the minimum is a $d_{ij}$, combine particles $i$ and $j$, and
  go back to step one.

\item Iterate until all particles have been declared jets.

\end {itemize}
This algorithm is usually what is meant by when the $k_T$ algorithm is
referred to, but you may occasionally see it referred to as the {\it
  inclusive} $k_T$ algorithm, as there is a related ({\it
  ``exclusive''}) variant \cite{Catani:1993hr}. Note that the
parameter $R$ functions as an angular cut-off: two particles separated
by a distance $R_{ij} > R$ will never be combined, regardless of the
$p_T$'s of the particles (this does not necessarily preclude both
particles being clustered into the same jet later).  In fact, with
this jet algorithm, arbitrarily soft particles can become
jets. Therefore jets are customarily returned down to some finite
$p_T$ cutoff, typically tens of GeV.

Because the $k_T$ algorithm clusters particles beginning with soft
particles and working its way up to harder particles, the algorithm
tends to construct irregular jets which depend on the detailed
distribution of soft particles in an event. For this reason, $k_T$
jets are not especially practical for hadron colliders: irregular jets
are hard to calibrate, and the jets are quite sensitive to unrelated
radiation in the event. 

Other sequential algorithms are obtained by using different metrics.
The {\it Cambridge-Aachen} or C-A algorithm is obtained by taking
\cite{Dokshitzer:1997in}
\beq 
d_{ij} = \frac{ \Delta R_{ij}^2}{R^2} , \phantom{space} d_{iB} = 1.
\eeq
This metric clusters particles based only on their angular separation,
giving a nicely geometric interpretation of jets.  The C-A algorithm
still reflects aspects of the QCD parton shower, in particular the
{\it angular ordering} of emissions.  However, it is less directly
related to the structure of QCD parton splitting functions than the
$k_T$ algorithm is, and represents a compromise between reflecting the
structure of the parton shower and maintaining some insensitivity to
soft radiation.

The {\it anti-$k_T$} algorithm entirely abandons the idea of mimicking
the parton shower \cite{Cacciari:2008gp}.  Here, the metric is
\beq 
d_{ij} = \min \left(\frac {1} {p_{T,i} ^ 2},\frac {1} {p_{T,j} ^ 2}\right)
              \frac{ \Delta R_{ij}^2}{R^2} , \phantom{space} d_{iB} = \frac {1} {p_{T,i} ^ 2}.
\eeq
With this metric, particles are clustered beginning with the {\it
  hardest} particles.  This means that the most energetic cores of
jets are found first. As soft particles clustered later have a minimal
impact on the larger four-momentum of the jet core, the anti-$k_T$
algorithm tends to cluster particles out to distances $R$ from the
core of a jet, yielding very regular jets. Anti-$k_T$ jets are
therefore much easier to calibrate at experiments, and the anti-$k_T$
algorithm has become the default used at the LHC.

Let us conclude this section by emphasizing that all sequential jet
algorithms return not only a list of jets but a {\it clustering
  sequence} for the event.  Varying the radial parameter $R $ simply
acts to move the resolution scale up and down the clustering sequence,
making it very easy to study how jet distributions and multiplicities
depend on the angular resolution $R$.  In particular, for the C-A
algorithm, the cluster sequence regarded as a function of $R$ has a
purely geometric interpretation as resolving the event on different
angular scales.

All three sequential jet algorithms discussed here also share the same
{\it reach}, that is, regardless of the chosen metric, a splitting
$P\to i j$ will not be combined if the angular distance between the
daughters exceeds the chosen jet radius, $\Delta R_{ij}> R$. This
means that, to leading order, perturbative computations of quantities
such as jet rates are identical between all three algorithms.

Finally, the infrared and collinear safety of all three sequential jet
algorithms can be easily checked by asking how the cluster sequence
would change with the addition of a soft or collinear emission.  For
the shower-sensitive $k_T$ and C-A metrics, infrared and collinear
safety follows automatically.  The anti-$k_T$ metric is also
manifestly IR- and collinear-safe, as can be seen with a little more
thought: anti-$k_T$ recombinations are clearly collinear-safe, since
collinear splittings are combined near the beginning of the sequence.
IR safety also follows, as soft radiation has negligible impact on the
jet built out from the hard core.

\subsection{Jets at the LHC}

The main subject of these lectures are the possibilities and uses of
jets to discover physics at and beyond the electroweak scale, which
means, for practical purposes, at the LHC.

It is important to remember that events at LHC are a busy hadronic
environment.  In addition to the showering and hadronizing hard partons
which we want to study, there are large amounts of soft, unassociated
radiation from (1) the {\it underlying event}, that is, the remnants
of the scattering protons; (2) possible {\it multiple interactions},
that is, additional collisions of partons arising from
the same $p$-$p$ collision as the hard interaction; and (3) {\it
  pile-up}, additional $p$-$p$ collisions from other protons in the
colliding bunches. These additional sources of radiation contribute a
potentially sizable and largely uniform backdrop of hadronic activity
that, when clustered into jets, will partially obscure the features of
the hard interaction that we would like to reconstruct.

The default jets used at the LHC are formed using the anti-$k_T$
algorithm, with cone sizes $R=0.4, 0.6$ (at ATLAS) and $R=0.5,0.7$ (at
CMS).  These specific choices of $R$ come from a compromise between
(1) the desire to collect all the radiation from a single parton, and
(2) the desire {\it not} to sweep up an excessive amount of unrelated
radiation.  

Many advances have combined to make jets at the LHC a particularly
fertile field.
\begin {itemize}

\item {\it advances in experiment}: the calorimeters at ATLAS and CMS
  have much finer resolution than in previous experiments, allowing a
  much more finely grained picture of events. Moreover, {\it local}
  calibration of jets allows jets to be considered on multiple
  scales.

\item {\it advances in computation}: the development of fast
  algorithms \cite{Cacciari:2005hq} allows broad implementation
  of sequential recombination.

\item {\it advances in energy}: the LHC center of mass energy is large
  enough that particles with weak scale masses (i.e., $Z, W, t,$ and
  $H$) will for the first time have an appreciable cross-section to be
  produced with enough of a boost to collimate the daughter partons.
  The simple picture that one parton corresponds to one jet breaks
  down badly in this case, and new tools are needed to separate out
  collimated perturbative decays from QCD showers.

\end {itemize}

There are several reasons to be interested in boosted particles.  Very
often, there is theoretical motivation to focus on a particular slice
of phase space where the daughter particles are necessarily
boosted. High mass resonances are the simplest such examples.  For
instance, a resonance $\rho_C$ with mass $m_\rho \gtrsim 1.5$ TeV which
decays to pairs of gauge bosons would yield highly boosted $VV$ pairs.
 
Even in the absence of a resonance or other mechanism to
preferentially populate boosted regions of phase space, looking for
boosted signals can also be useful for improving the signal to
background ratio.  Changing the reconstruction method changes what the
experimental definition of the signal is, and therefore necessarily
the backgrounds change as well. This can sometimes---but not
always!---be enough of an advantage to make up for the reduction in
signal rate that comes from selecting only the boosted region of phase
space.  Background reduction comes in two forms.  In high multiplicity
final states, combinatoric background is often prohibitive.  When some
or all of the final state particles are boosted, the combinatoric
background is greatly reduced.  But it is also possible to use boosted
selection techniques to identify regions where the background from
other physics processes is intrinsically reduced.

To appreciate the need for new reconstruction techniques at the LHC,
consider the production of top quarks at fixed center of mass energy
$\sqrt{\hat s}$.  Choosing some angular scale $R_0$, we can ask, what
fraction of top quarks have all three, only two, or none of their
partonic daughters isolated from the others at the scale $R_0$?  This
gives a zeroth order estimate of how well a jet algorithm with $R=R_0$
will be able to reconstruct the three partonic top daughters as
separate jets.  The answer we get depends sensitively on both $R_0$
and $\sqrt{\hat s}$:

%%%%%%%%%%%%%%%%
\begin{table}[h]
\begin{center}
\begin{tabular}{ccccc}
\hline \hline

 $\sqrt{\hat s}$ & $R_0$          & 3  & 2  & 1 \\
\hline
 1.5 TeV & $0.4$           & 0.55   & 0.45   & ---   \\
 1.5 TeV & $0.6$           & 0.2   & 0.6   & 0.2   \\
 2.0 TeV & $ 0.6$           & 0.1   & 0.45   & 0.45   \\

\hline \hline
\end{tabular}
\caption{Resolved parton multiplicities in $t\bar t$ events}
\end{center}
\end{table}
%%%%%%%%%%%%%%%%%%%%%%%%

Clearly, tops produced in the very interesting super-TeV regime
$\sqrt{\hat s}\gtrsim$ TeV straddle the borderlines between several
different topologies.  It would be much more desirable to have a
flexible reconstruction method that could handle semi-collimated tops
in a unified way.

To see how we can go about building such reconstruction techniques,
let's start by considering one of the landmark jet substructure
analyses: the case of a boosted Higgs decaying into $ b \bar b$.

%%%%%%%%%%%%%%%%%%%%%%%%%%%%%%%%%%%%%
\subsection {Boosted Higgs}
%%%%%%%%%%%%%%%%%%%%%%%%%%%%%%%%%%%%%

This analysis will introduce us to several ideas that will be
important tools in our boosted analysis toolbox: fat jets, jet mass,
jet grooming, and sequential de-clustering.

Searching for the Higgs in its decay to $b\bar b$ is very difficult at
the LHC, due to overwhelming QCD backgrounds.  Even in associated
production, $pp\to H Z, HW$, the background processes $Z+b\bar b$,
$W+b\bar b$, and even $t\bar t$ are overwhelming.  Nonetheless, thanks
to Ref.~\cite{Butterworth:2008iy}, $pp\to H V,\, H\to b\bar b$ is now
an active search channel at the LHC.

To be specific, let's consider the process $p p \to H Z$, followed by
$ H\to b\bar b$, and $Z\to \ell^+\ell^-$.  The traditional approach to
this signal would be to look for final states with a leptonic $Z$ and
2 $b$-tagged jets, construct the invariant mass of the jets, and look
for a peak in the distribution of $m_{b\bar b}$.  The new approach is
instead to focus on events where the Higgs is produced with
substantial $p_T$, $p_{T,H} > 200$ GeV, and cluster these events with
a large ($R= 1.2$) jet radius, such that all of the Higgs decay
products are swept up in a single fat jet.  The signal is now a
leptonic $Z$ + a fat ``Higgs-like'' jet, and the background to this
signal is now $Z+$ one fat jet rather than $Z+b\bar b$.  What we'll
see is that jet substructure offers us enough quantitative precision
in what we mean by a ``Higgs-like'' jet to reduce the background by an
extent that makes up for the acceptance price demanded by the high
$p_T$ cut.

For an unboosted search, the ultimate discriminator between signal and
background is the $b$-$\bar b$ invariant mass: to find a resonance,
look for a bump in the $b$-$\bar b$ mass spectrum.  Now that we have
boosted the Higgs and collected it into a single fat jet, the Higgs
mass should be reflected in the invariant mass of the fat jet itself.
To understand jet masses for the background, let's take a quick look
at how jet masses are generated in QCD.  

\paragraph{Jet Mass}.  Partons are generally massless (we
will neglect the $b$ quark mass), but jets are not.  Jet mass in QCD
arises from emission during the parton shower, and as such we can
calculate the leading contribution.  Jet mass, like most perturbative
jet properties in QCD, is dominated by the first emission.  Let's
consider for concreteness a quark emitting a gluon, and work in the
collinear regime (small $R$).  In this approximation, we can consider
the jet in isolation from the rest of the event, neglecting
interference and splash-in, and we can approximate the QCD splitting
functions with the singular portions.  Doing so,
the amplitude to radiate an extra parton can be written as
\beq
d\sigma_{n +1} \approx d\sigma_n \, dz \, \frac{dt}{t} \, \frac {\alpha_s}{2\pi} \, {\mathcal P} (z),
\eeq
where $t$ is the virtuality of the parent $P$, $z = E_q/E_P$ is the
fraction of the parent energy retained by the daughter quark, and the
splitting function ${\mathcal P}(z)$ for $q\to q g$ is given by
\beq
{\mathcal P}(z) = C_F\frac {1+z^2}{1-z} .
\eeq
The parent virtuality $t$ is of course the jet mass-squared.  In the
collinear limit,
\beq
t = E_P^2 z (1-z) \theta^2 = (p_{T,P} \cosh \eta)^2 z(1-z)\theta^ 2.
\eeq
Integrating over rapidity, we can approximate the average jet
mass-squared as:
\beq
\langle m^2 \rangle \approx p_{T,P}^2 \int_0^{R^2} \frac{d\theta^2}{\theta^2} \, \int dz \,
                 z (1-z) \theta^2 \frac {\alpha_s}{2\pi} {\mathcal P}(z).
\eeq
Note the limits on the $\theta$ integral: this is where the choice of
jet algorithm enters.  As established above, for all sequential jet
algorithms, only radiation at angles smaller than $R$ will be
clustered into the jet.  Strictly, we should use a running $\alpha_s$
evaluated at a scale set by the relative transverse momentum of the
splitting, but to get a quick estimate, let's perform the integral in
the approximation that $\alpha_s$ is constant.  We then obtain
\beq
\langle m^2 \rangle \approx \frac{\alpha_s}{\pi} \, \frac{3}{8} C_F \, p_{T}^2 R^2 .
\eeq
The jet mass scales like $p_T$, as it had to, and is suppressed by
$(\alpha_s/\pi)^{1/2}$.  To this order the mass increases linearly
with $R$. The exact value of the numerical coefficient will in general
depend on the quark versus gluon content of the jet sample.  For
instance, the major QCD background for a doubly $b$-tagged boosted
Higgs comes from the splittings $g\to b\bar b$, where the splitting
function is
\beq
{\mathcal P}(z) = C_A (z^2+(1-z)^2) ,
\eeq
giving, in the constant-$\alpha_s$ approximation,
\beq
\langle m^2 \rangle \approx \frac{\alpha_s}{\pi} \, \frac{1}{20} C_A \, p_{T}^2 R^2 .
\eeq

\paragraph{Coming back to the Higgs,} consider now a
splitting $P\to i j$.  We have $m^2\approx 2 p_i\cdot p_j \sim
 p_{T,i}p_{T,j} \Delta R_{ij}^2 =  z (1-z)
p_{T,P}^2 \Delta R_{ij}^2$.  In other words, just from kinematics we
can express the opening angle in terms of the parent mass and $p_T$:
\beq
\label{eq:Rofboost}
\Delta R_{ij} \sim \frac {m}{p_T} \frac { 1} {\sqrt{ z(1-z)}} \sim \frac{2 m}{ p_T} .
\eeq
Now consider the $k_T$ metric evaluated on this splitting $P\to i j$:
\beq
y_{ij} = \min (E_{T,i} ^ 2, E_{T, j} ^ 2)\Delta R_{ij}^2 = p_T^2 z^2\Delta R_{ij}^2
        \approx \frac {z} {1-z} m^ 2 . 
\eeq
For jets with a fixed mass $m$, cutting on the splitting scale
$y_{ij}$ then can separate QCD jets, which have a soft singularity
$\propto 1/z$, from boosted Higgses, which have a flat distribution in
$z$.  \footnote{This is a little quick: not all QCD splitting
  functions have a soft singularity, and in particular $g\to q \bar q$
  does not. However, $P_{g\to q \bar q}(z)$ is not flat in $z$, and in
  particular is minimized at the symmetric value $z=1/2$, so cutting
  on $y_{ij}$ can still help suppress this background.}

Moreover, a boosted Higgs will go from a mass $m_H$ to massless
daughters in one step, while QCD splittings prefer to shed virtuality
gradually.  To see this, consider the Sudakov form factor, which
exponentiates the splitting functions to obtain the probability of
evolving from an initial virtuality $t_0$ to a final virtuality $t$
without branching:
\beq
\Delta (t) = \exp \left[-\int_{t_0} ^ t \frac{dt'}{t'}\, dz
  \frac{\alpha_s}{2\pi}{\mathcal P}(z) \right] .
\eeq
Evaluating $\alpha_s = \alpha_s (t')$ and using an IR cut-off to
regulate the splitting functions, at large $t$, one can work out that
\cite{QCDnCP}
\beq
\Delta (t) \propto \left(\frac{t_0}{t}\right)^p
\eeq
for an exponent $p>0$, in other words, $\Delta (t)\to 0$ for large
$t$.  In other words, the probability of a QCD jet making a large jump
in mass at a branching falls off as $m^{-2p}$.\footnote{In fact,
  taking higher order corrections into account, one finds that the
  Sudakov form factor goes to zero even faster than polynomially for
  large $t$.}

We have now identified two ways in which a Higgs boson $H$ decaying
perturbatively to $b\bar b$ will behave very differently from a QCD
parton branching: the splitting will be symmetric, and show a sudden
drop in parton mass.  The search algorithm for finding a boosted Higgs
looks for a splitting inside the Higgs jet that behaves like a
perturbative decay, and works as follows:
\begin {itemize}
\item Cluster the event on a large angular scale
  (Ref.~\cite{Butterworth:2008iy} uses $R=1.2$), using the C-A
  algorithm.  Large angular scales are necessary in order to get good
  acceptance for collecting both Higgs decay products into a single fat jet: from
  Eq.~\ref{eq:Rofboost}, we can see that the $b$-$\bar b$ separation
  for a 125 GeV Higgs boson is $R_{b\bar b}\lesssim 1$ for $p_T\gtrsim 200$
  GeV.  We choose the C-A algorithm because it is a good compromise
  between accurately reflecting the shower structure of QCD, and
  minimizing sensitivity to soft  radiation in the event.
 
\item Now, given a hard fat jet, successively {\it unwind} the jet by
  undoing the cluster sequence one branching at a time.  At each
  branching $P\to i j$, check to see whether the splitting looks
  sufficiently non-QCD-like, by asking that the branching be both {\it
    hard},
\beq
\max (m_i, m_j) < \mu m_P
\eeq
for some parameter $\mu$, and {\it symmetric},
\beq
y_{ij} > y_{cut}
\eeq
for some choice of $y_{cut}$.

\item If the splitting fails to be sufficiently hard and symmetric,
  discard the softer of $i$ and $j$, and continue to unwind the
  harder.

\item Continue until either an interesting splitting has been found
  or you run out of jet.

\end {itemize}
This procedure, often referred to as the ``splitting'' or
``mass-drop'' procedure, identifies an interesting Higgs-like
splitting $H\to b\bar b$, which determines a characteristic angular
scale $R_{b\bar b}$ for a particular event. Once this scale $R_{b\bar
  b}$ has been identified, we benefit greatly by using smaller scales
to resolve the event, rather than the large $R=1.2$ scale we started
with.

The reason is the following: starting with such a large jet, we are
guaranteed to sweep up a large amount of unassociated radiation along
with the Higgs decay products. The effect of this unassociated
radiation is to smear out the mass resolution.  The invariant mass is
especially vulnerable to distortion from even soft unassociated
radiation, because evaluating $m^2 = E^2 -\vec{p}^2$ depends on large
cancellations. The amount of distortion scales like 
\beq
\frac{d\langle m^2\rangle}{dR} \approx \Lambda_{soft} p_{T,J} R^3,
\eeq
in the approximation that unassociated radiation contributes a
constant energy $\Lambda_{soft}$ per unit rapidity: the jet area
scales like $R^2$, while the incremental contribution to the invariant
mass from a soft particle at distance $R/2$ from the jet core
contributes as $\Lambda p_{T,J} R/2$ \cite{Dasgupta:2007wa}.  So to
recover mass resolution, it is vital to whittle down our initial fat
jet to jets only as big as necessary to capture the radiation from the
Higgs decay products.

In fact, we have already started whittling.  The splitting procedure
discards soft, wide-angle radiation clustered into the jet on its way
towards finding the Higgs-like splitting. This by itself helps to
clean up the mass resolution. But we can do better: given the scale
$R_{b\bar b}$ which is our best guess at the angular separation of the
Higgs' daughter particles, we can resolve the fat jet at the {\it
  filtering} scale $R_{filt} = \min (R_{b\bar b}/2, 0.3)$, and keep
only the three hardest subjets.  We keep three, rather than two,
subjets in order to capture final-state radiation off of one of the
$b$ quarks.

\begin{figure}
\begin{center}
\includegraphics[width=2.75in]{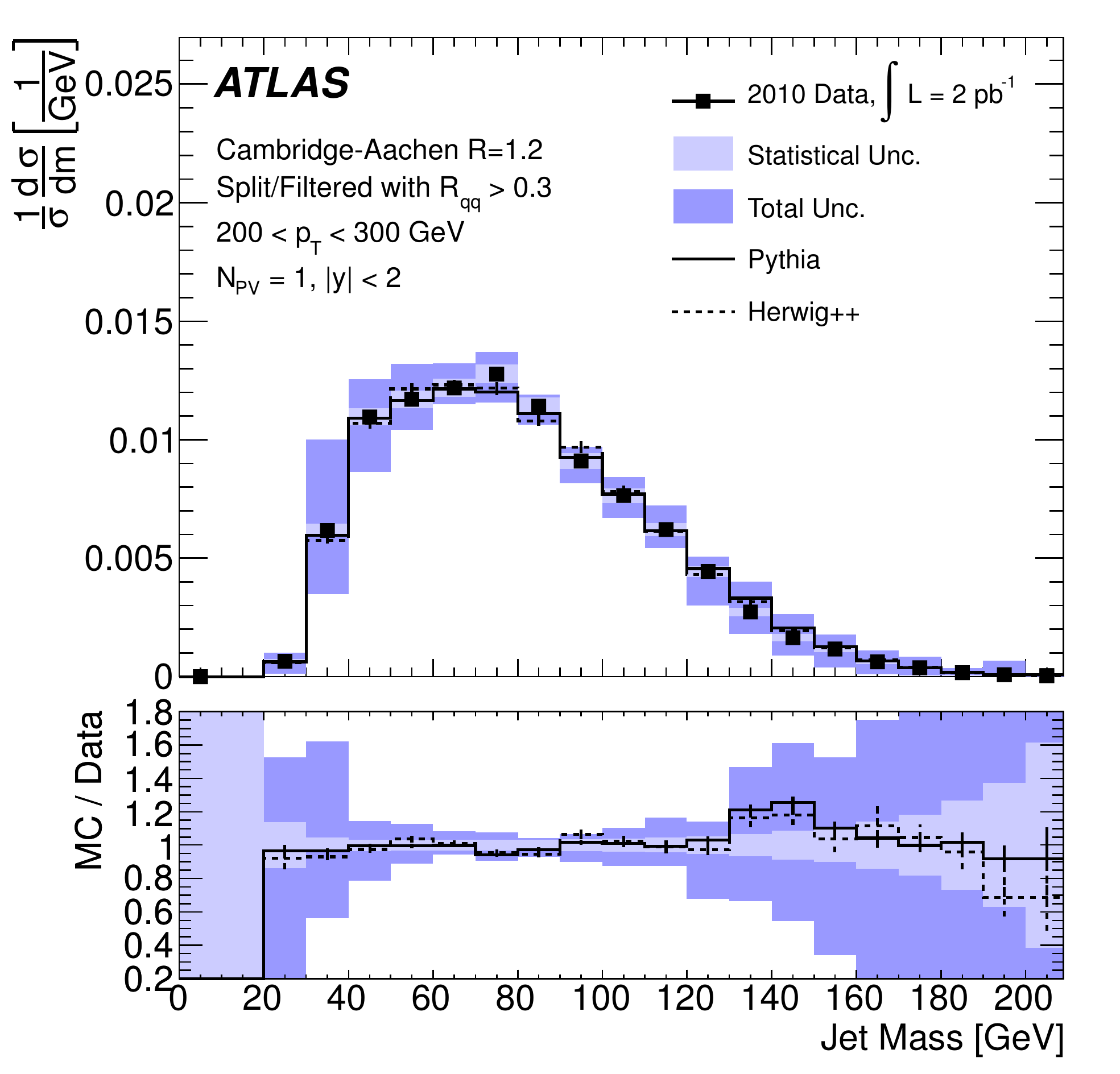}
\includegraphics[width=2.75in]{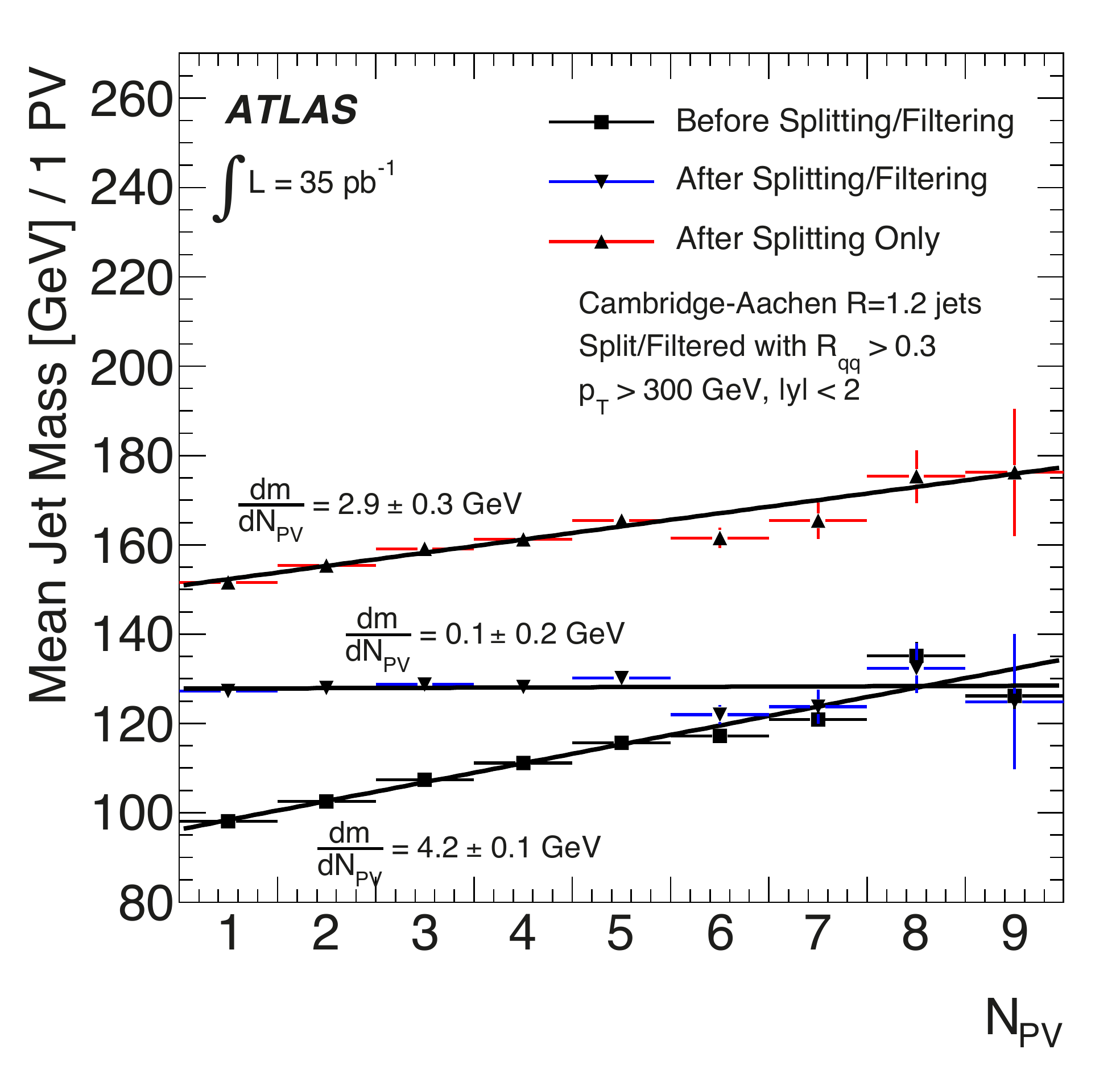}
\caption{(Left) The distribution of jet mass for fat C-A jets (after
  splitting and filtering).  Note the reasonable agreement between
  data and predictions from two different shower MCs. (Right) Average
  jet mass as a function of the number of primary vertices
  $N_{PV}$. Note that after filtering, the jet mass has little to no
  dependence on $N_{PV}$.  From Ref.~\cite{ATLAS:2012am}.}
\label{fig:atlasdata}
\end{center}
\end{figure}

Finally, demanding that the two hardest filtered subjets be
$b$-tagged, Ref.~\cite{Butterworth:2008iy} finds that the Higgs can
be seen in this channel with $5\sigma$ significance in 30 fb$^{-1}$
(at 14 TeV, combining $Z\to \ell^+ \ell^-$, $Z\to \nu\bar \nu$, and
$W\to \ell\nu$), and signal-to-background of ${\mathcal O}(1).$
However, we emphasize that {\it this and all other LHC
  phenomenological studies are based on expectations from Monte
  Carlo}.  Even very sophisticated Monte Carlos necessarily capture
only an approximation to the full physics of QCD. For this reason,
both validation in data on one hand and formal theoretical study on
the other are critical.  Let us then end this section by showing a
couple of the most important early experimental results.  In
Fig.~\ref{fig:atlasdata}, we show two plots from
Ref. \cite{ATLAS:2012am}.
%%%%%%%%%%%%%%%%%%%%%%%%%%%%%%%
%%%%%%%%%%%%%%%%%%%%%%%%%%%%%%%
On the left, we see that shower Monte Carlos do a reasonable job of
predicting the spectrum of jet masses for the QCD background.  On the
right, the jet mass is plotted as a function of the number of primary
vertices $N_{PV}$ in an event, or in other words, the amount of
pileup.  Note that after filtering, the jet mass has little to no
dependence on $N_{PV}$, indicating that filtering is successfully
isolating the hard process.  Note also that filtering is {\it
  necessary}: prior to filtering, the dependence of jet mass on
$N_{PV}$ is significant, and in the 2012 operating environment average
pileup multiplicity is $N_{PV}\gtrsim 30$.

Heartened by this evidence that our theoretical techniques have a
reasonable relationship with reality, we will proceed in the next
section to discuss more ideas for cleaning up pileup, and more jet
properties which can discriminate signals from QCD backgrounds.
 
\section{Lecture II: Jet Grooming and Jet Shapes}

Our last section concluded with a walk-through of the pioneering
boosted Higgs study, where we saw examples of two topics we will be
discussing in this lecture, namely jet grooming and jet shapes.

\subsection{Jet grooming}

In the boosted Higgs analysis discussed in the previous lecture, we
saw that jet mass resolution was badly degraded by the presence of
unassociated radiation in the jet, and introduced the process of {\it
  filtering} to mitigate these contributions.  Filtering is one of
several {\it jet grooming} algorithms, all of which are designed to
``clean up'' jets by subtracting the contributions of unassociated
radiation.

\paragraph{Trimming}\cite{Krohn:2009th}, similarly to filtering,
reclusters the constituents of a fat jet and retains a subset of the
subjets, but has a different criterion for keeping subjets. For each
jet of interest, the algorithm is:
\begin {itemize}

\item Recluster the constituents using some jet algorithm (the
  original reference specifies $k_T$), and resolve on a fixed small
  angular scale $R_0$.

\item Keep each subjet $i$ that passes a $p_T$ threshhold,
\beq
p_{T,i} > f \Lambda_{hard}
\eeq
 for a cutoff parameter $f_{cut}$ and a hard momentum scale $\Lambda$.

\item  The final trimmed jet is the sum of the retained subjets.

\end {itemize}
The essential idea is that radiation we want to keep tends to be
distributed in clusters, reflecting a parent parton emission, while
unassociated radiation we don't want to keep is more uniformly
distributed. Asking that radiation cluster sufficiently on small
scales then preferentially picks out the radiation which ultimately
originated from a parent hard parton.  The $k_T$ algorithm was
originally proposed here because it increases the chances that soft
FSR will be kept: since clustering in the $k_T$ metric works from soft
up, using $k_T$ increases the chance that a relatively soft parton
emitted in the parton shower will be reconstructed and pass above the
$p_T$ threshold. But it is possible to imagine using other algorithms
for the small-scale reclustering, and indeed implementations using C-A
\cite{BOOST2010} or even anti-$k_T$ \cite{Falkowski:2010hi} have been
seen to be effective.

The trimming algorithm is simple to state; the detailed questions
arise when we ask how the parameters should be chosen, and in any
particular application parameter choices should be optimized for the
specific process under consideration.  Typical values for the small
angular scale range between $0.2\leq R_0 \leq 0.35$; for $R_0$ much
smaller than $R_{min}=0.2$, the finite angular resolution of the
calorimeter starts to introduce irregularities.  Good choices for
$\Lambda$ are either the total jet $p_{T}$, for dijet events or other
such events where all jets have similar $p_T$s, or the scalar sum
transverse energy of the event, $H_T$, if jets have some spread over a
broader range of $p_T$s.  Typical values for the cutoff parameter
$f_{cut}$ range between $10^{-2}$ (more typically for jet $p_{T}$) and
$10^{-3}$ (for event $H_T$): this tends to work out to keeping subjets
down to a 5 to 10 GeV threshold.

\paragraph{Pruning} \cite{Ellis:2009su,Ellis:2009me} builds
on the observation that the mass-drop algorithm improves mass
resolution on boosted hard decays even before the filtering step, by
discarding soft wide-angle radiation clustered into the fat jet at the
final stages.  In the C-A algorithm, the typical last clusterings in
the fat jet are of stray soft radiation, usually unassociated with the
parent particle, at wide angles to the jet core. These late,
wide-angle clusterings have a disproportionate effect on jet mass.

Pruning adapts the splitting algorithm to specifically check for soft,
wide-angle splittings, and throw them away. The algorithm is:
\begin {itemize}

\item Given a jet $J$, recluster its constituents with C-A, and then
  sequentially unwind the cluster sequence.

\item At each splitting $P\to ij$, check whether the splitting is both
  {\it soft},
\begin{equation}
z = \frac{\min (p_{T,i},p_{T,j})}{p_{T,P}} < z_{cut} ,
\end{equation}
and at {\it wide angle},
\begin{equation}
\Delta R_{ij} > D_{cut} .
\end{equation}
If so, then drop the softer of $i$, $j$, and continue unwinding the harder.

\item  Stop when you find a sufficiently hard (or collinear) splitting.

\end {itemize}
Again, this algorithm has parameters that must be optimized
specifically for each process under consideration.  Typical values of
$z_{cut}$ are $z_{cut}\approx 0.1$, while the radial separation should
be tuned to the expected opening angle for a hard process, $D_{cut}
\approx 2 m/p_T \times 1/2$.

\begin{figure}
\begin{center}
\includegraphics[width=2.75in]{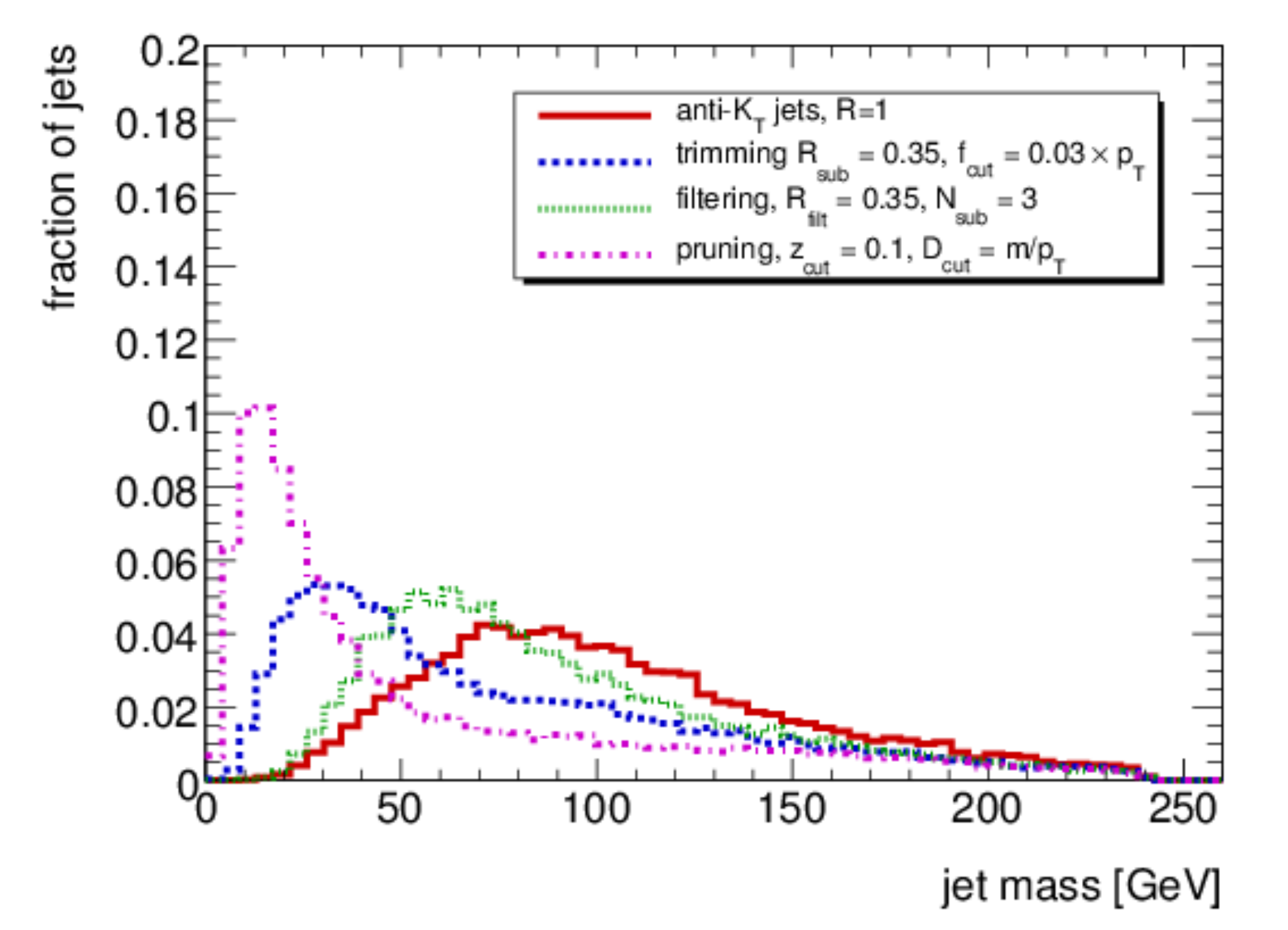}
\includegraphics[width=2.75in]{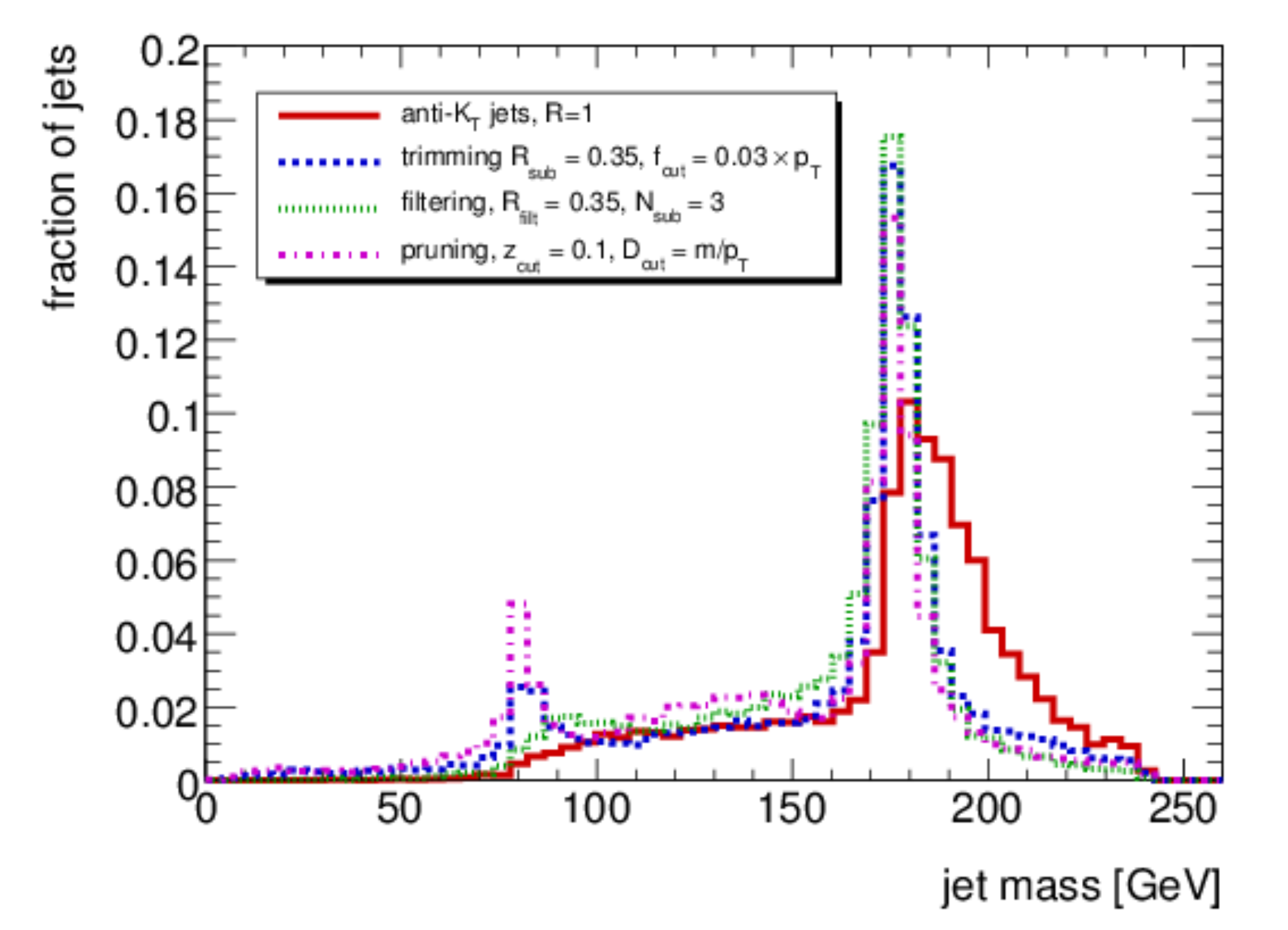}
\caption{The operation of filtering (green, dotted), trimming (blue,
  dashed), and pruning (purple, dash-dotted) on background QCD jets
  (left) and boosted top jets (right).  From
  Ref.~\cite{BOOST2010}.}
\label{fig:grooming}
\end{center}
\end{figure}

\paragraph{Grooming in action.} All three grooming
techniques (filtering, trimming, and pruning) improve signal to
background by both improving mass resolution for signal {\it and}
suppressing QCD background.  QCD jets, whose jet masses are generated
by relatively softer and less symmetric emissions, are more likely to
have their masses shifted substantially downward by jet grooming than
collimated perturbatively decaying particles are, thus depleting the
background to high-mass searches.  Both the sharp gain in signal
mass resolution and the depletion of the high mass background can be
seen in Fig.~\ref{fig:grooming}.

%%%%%%%%%%%%%%%%%%%%%%%%%%%%%%%
%%%%%%%%%%%%%%%%%%%%%%%%%%%%%%%

We can also see in Fig.~\ref{fig:grooming} that the different grooming
techniques all act slightly differently on background massive QCD
jets\cite{BOOST2010}.  QCD jets with high masses dominantly have this
mass generated by a relatively hard perturbative emission, which all
algorithms are designed to retain, so performance between the
different algorithms is similar.  However, the effects of the
different grooming algorithms on QCD backgrounds are still
sufficiently distinct that some benefit can be obtained in applying
multiple grooming algorithms \cite{Soper:2010xk}.

At low masses, the differences between the grooming algorithms become
more pronounced.  QCD jets at low masses are dominated by a hard core.
Filtering keeps a fixed number $N=3$ of subjets, and therefore retains
relatively soft radiation.  Trimming, by contrast, will typically drop
all radiation except that within $R_{sub}$ of the jet core.  Pruning
will also typically drop all but the radiation in the core, but the
resolution radius $D$ is set to scale like $ m/p_T$, and therefore
$D\to 0$ as $m\to 0$.  Thus at small masses typically
$R_{prune}<R_{sub}$, so pruning acts more aggressively than trimming.

%%%%%%%%%%%%%%%%%%%%%%%%%%% 
\subsection{Jet Shapes}
%%%%%%%%%%%%%%%%%%%%%%%%%%%

Another feature of the boosted Higgs analysis we saw in the previous
lecture was the importance of {\it jet mass}, which allowed us to
concentrate signal in a sharp peak on top of a falling background
\cite{Skiba:2007fw}.  Jet mass is an example of a {\it jet shape}: a
function $f$ defined on a jet $J$ that quantifies the properties of
the jet without the (explicit) use of any jet algorithm.  The approach
is conceptually akin to {\it event shapes}, which allow quantitative
study of QCD without requiring specific characterization of an event
in terms of jets, and indeed many jet shapes are descendants of event
shapes.

Before discussing individual jet shapes, let us make two general
comments.  First, as we saw for jet mass, jet shapes are vulnerable to
the inclusion of unassociated radiation, particularly pile-up, into
jets, to a greater or lesser extent depending on the particular jet
shape, and the sensitivity of the jet shape to unassociated radiation
can be important.  Second, one should bear in mind that any reasonable
jet shape needs to be both infrared- and collinear-safe.  Any linear
function of particles' $p_T$ is automatically safe; factorization
 theorems for other jet shapes can be proven \cite{Ellis:2010rwa}.

\subsubsection{Radial distribution of particles within a jet}
\label{sec:raddist}

The probability of a showering parton to emit a daughter parton
depends on the running coupling $\alpha_s$ evaluated at the $k_\perp$
scale of the splitting.  Jet shapes which measure the angular
distribution of particles in an event are therefore measuring both the
strength and the running of the strong coupling constant, and are
classic probes of QCD.  These jet shapes are also sensitive to the
color charge of the parent parton: since $C_F < C_A$, an initial gluon
will radiate more, and at wider angles, than an initial quark.

\paragraph{Jet Broadening} is a classic $e^+ e^-$
observable.  Given a thrust axis $\hat n$, we can partition the
particles $i$ in an event into hemispheres according to
$\mathrm{sign}(\vec{p}_i\cdot\hat n )$, which for dijet-like events is
equivalent to associating each particle to a jet.  Hemisphere
broadening is then defined as the momentum-weighted transverse
spread of the particles,
\beq
B_H = \frac{1}{\sum_{i\in H} |\vec{p}_i|} \sum_{i\in H} |\vec{p}_i \times \hat n |
\eeq
where the sum runs over all particles $i$ in a hemisphere $H $.

\paragraph{Differential and Integrated Jet Shapes} are,
thanks to a historic quirk of nomenclature, names for two specific jet
shapes: the so-called {\it differential jet shape} $\rho(r)$ and the
{\it integrated jet shape} $\Psi (r)$, which characterize the radial
distribution of radiation inside a jet.  These jet shapes are also
sometimes called the {\it jet profile}.  Both of these shapes are
defined on an ensemble of $N$ jets formed with radius $R$.  Then for
$r < R$, the integrated jet shape $\Psi(r)$ is the ensemble average of
the fraction of a jet's $p_T$ which is contained within a radius $r$
from the jet axis.  Defining $r_i$ as the distance of a constituent
$i$ from the jet axis,
\beq
\Psi (r)  = \frac{1}{N} \sum_{J} \sum_{i\in J} \frac{p_T(0 < r_i <r)}{p_{T, J}}.
\eeq
Here the second sum runs over all constituents $i$ of a jet $J$.
The differential jet shape $\rho(r)$ is then given by
\beq
\rho (r) =\frac {1} {\delta r}\, \frac {1} {N}\sum_{J} \sum_{i\in J} \frac{p_T(r <r_i < r+\delta r)}{p_{T, J}} .
\eeq
These variables are often included in the suite of QCD precision
measurements performed by experimental collaborations, as for instance
in the ATLAS study \cite{Aad:2011kq}, and are useful for validating
parton shower models.

\paragraph{Girth} is another jet shape which probes the
radial distribution of radiation inside a jet.  Let $r_i$ again
be the distance between a constituent $i$ and the jet axis. Then the
girth of a jet $g_J$ is the linear radial moment of the jet,
\beq
g_J = \sum_{i\in J} \frac {p_{T,i}r_i} {p_{T,J}}.
\eeq
In the collinear limit $\theta\to 0$, girth becomes equivalent to jet
broadening (where the thrust axis is replaced by the jet axis).  Girth
has been shown to be particularly useful for distinguishing
quark-initiated jets from gluon-initiated jets
\cite{Gallicchio:2011xq}.

\paragraph{Angularities} \cite{Almeida:2008yp} are a related
family of jet shapes, defined as a function of the parameter $a$:
\beq
\tau_a = \frac{1}{2E_J} \sum_{i\in J} p_{\perp,i} e^{(a-1)\eta_i} .
\eeq
Here $\eta_i$ is the separation in rapidity only between particle $i$
and the jet axis, and $p_{\perp,i}$ the momentum transverse to the jet
axis.

\subsubsection{Discriminating boosted decay kinematics}
\label{sec:shapesub}

The radial distribution jet shapes discussed in the previous section
are geared toward probing the characteristic shower structure of QCD.
Here we will discuss several examples of jet shapes which target evidence of
non-QCD-like substructure in jets.

\paragraph{Planar flow} \cite{Almeida:2008yp} considers the
spread of the jet's radiation in the plane transverse to the jet axis
(see also the closely related {\it jet transverse sphericity} shape
\cite{Thaler:2008ju}).  Since QCD showers are angular-ordered,
radiation subsequent to the first emission $P\to i j$ tends to be
concentrated between the clusters of energy defined by $i$ and $j$,
leading to a roughly linear distribution of energy in the jet. By
contrast, boosted three-body decays, such as boosted tops, have a more
planar distribution of energy.

Define the tensor
\beq
I^{ab} = \frac{1}{m_J}\sum_{i\in J} \frac{p^a_{i,\perp} p^b_{i,\perp}}{E_i},
\eeq
where the indices $a,\,b$ span the plane perpendicular to the jet
axis, and $\vec{p}_{ i,\perp}$ denotes the projection of particle
$i$'s momentum into this plane.  Letting $\lambda_1,\,\lambda_2 $ be
the eigenvalues of $I^{ab}$, the planar flow of a jet is given by
\beq
Pf_J=\frac{4\lambda_1\lambda_2}{ (\lambda_1+\lambda_2) ^ 2} =\frac {\det I}{(\tr I)^2}.
\eeq
With this normalization, $Pf_J\in (0,1)$. Monte Carlo studies have
demonstrated that QCD events do indeed peak at low values of $Pf$,
while boosted top decays show a relatively flat distribution in $Pf$,
but preliminary results show some sensitivity to shower modeling
\cite{Thaler:2008ju} and the utility of this shape in data is so far
unclear.

Note that neither $I^{ab}$ nor its eigenvalues are invariant under
longitudinal boosts.  For fully reconstructible events this is not a
worry in theory, as all events can be considered in the reconstructed
CM frame, but finite experimental resolution can become an issue in
transforming from the lab frame into the CM frame.

\paragraph{Template overlaps} define jet shapes based on
(aspects of) the matrix elements for boosted object decays
\cite{Almeida:2010pa}.  For example, consider the three body top quark
decay with intermediate on-shell $W$. The phase space for this decay
is (in the narrow-width approximation) determined by four parameters,
which can be parameterized as the solid angle governing the two-body
decays of both the $t$ and its daughter $W$.  Note that (1) the
azimuthal angle $\phi_t$ is meaningful, as the detector geometry is
not invariant under rotations around the top direction of motion, and
(2) this phase space has both $m_t$ and $m_W$ built in. A series of
templates describing this phase space can be generated by discretizing
the four-dimensional space.  To use these templates on a jet, the
method of template overlaps finds the template which has best overlap
with the kinematic configuration of the jet constituents according to
a chosen metric.  The ultimate variable is the numerical value of the
best overlap, which distinguishes between QCD jets and boosted tops.

\paragraph{$N$-subjettiness} \cite{Thaler:2010tr} takes a
different and more general approach to probing jet substructure via
jet shapes.  Given $N$ axes $\hat n_k$, we define $N$-subjettiness as
\beq
\tau_N =\frac{\sum_{i\in J}p_{T,i} \min (\Delta R_{i k} ) }{\sum_{i\in J} p_{T,i} R_0} 
\eeq
where $R_0$ is the jet radius, and $\Delta R_{ik}$ is the distance
between particle $i$ and axis $\hat n_k$. The smaller $\tau_N$ is, the
more radiation is clustered around the chosen axes, or in other words,
smaller values of $\tau_N$ indicate a better characterization of the
jet $J$ as having $N$ (or fewer) subjets.  Conversely, if $\tau_N$ is
large, then a description in terms of $>N$ subjets is better.

However, as QCD alone will happily make jets with subjets, to
differentiate boosted objects we need to probe not just the possible
existence of subjets, but their structure.  The real distinguishing
power of $N$-subjettiness occurs when looking at {\it ratios}. For
instance, a two-prong boosted particle such as a Higgs or $W $ will
have large $\tau_1$ and small $\tau_2$.  QCD jets which have small
$\tau_2$ will generically have smaller $\tau_1$ than for signal, as
the QCD jets are more hierarchical; conversely, QCD jets which have
large $\tau_1$ are generally diffuse, and will have larger $\tau_2$ as
well than for signal.  Thus the best single discriminating variable is
$\tau_2/\tau_1$, or, more generally
\beq
r_{N} = \frac{\tau_N}{\tau_{N-1}}
\eeq
for a boosted $N$-prong particle.

The question of how to determine the input subjet axes $\hat n_k$ is
an interesting one.  One approach, which is fast and perfectly
serviceable for most applications, is to use a jet algorithm, such as
exclusive $k_T$, to determine subjet axes.  Naturally, the results
then retain some dependence on the choice of jet algorithm used to
find the axes. Another approach is to marginalize over all possible
choices of $\hat n_k$, and choose the set which minimizes $\tau_N$
\cite{Thaler:2011gf}.  While this choice is computationally more
intensive, it removes the dependence on the jet algorithm choice, and
additionally guarantees the nice property that
\beq
\tau_{N-1} > \tau_{N} ,
\eeq
which holds only approximately if fixed subjet axes are used.

$N$-subjettiness is a conceptual descendent of the event shape
$N$-jettiness \cite{Stewart:2010tn}, which classifies events as being
$N$-jet-like without reference to jet algorithms.

\subsubsection{Color flow variables}

\begin{figure}
\begin{center}
\includegraphics[width=2.75in]{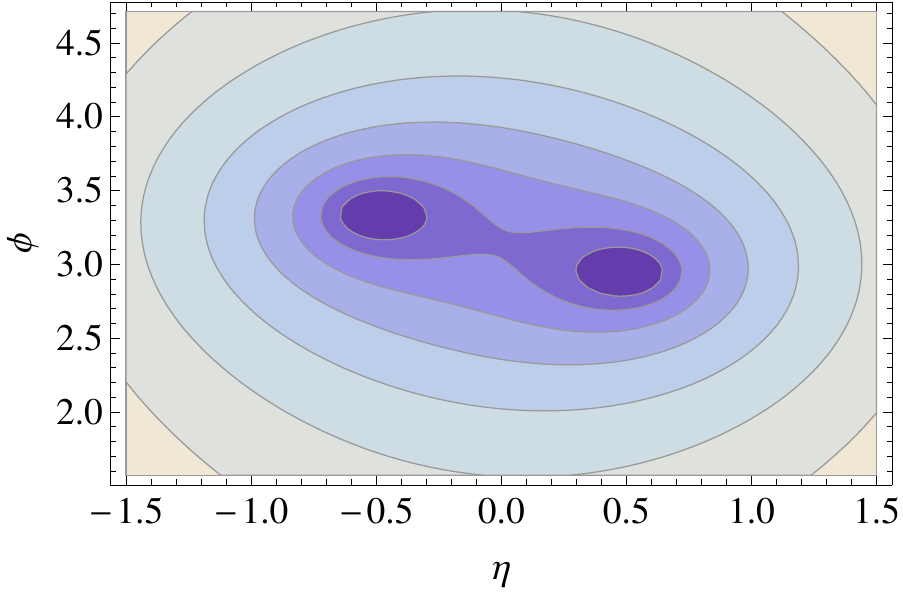}
\includegraphics[width=2.75in]{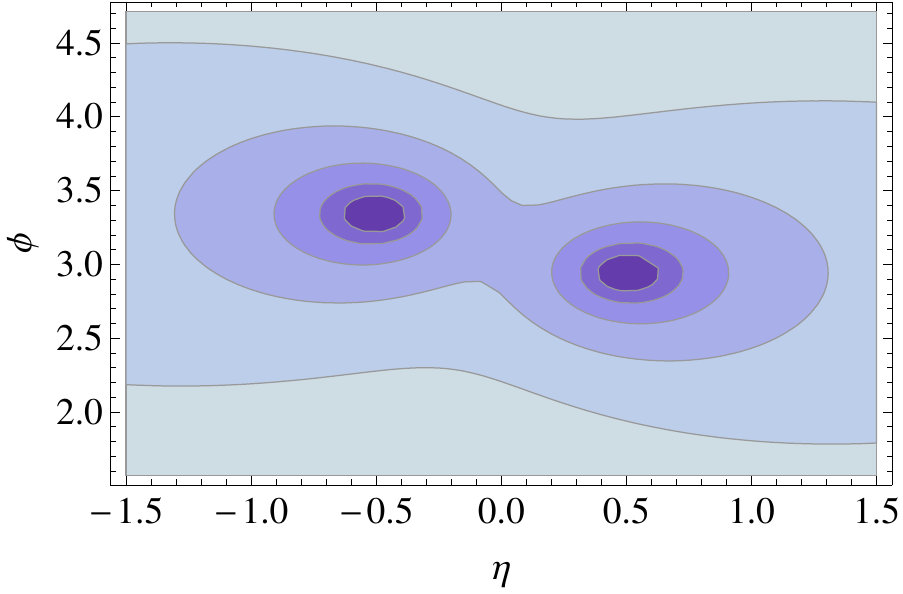}
\caption{Radiation patterns in the eikonal approximation for two
  triplet color sources color-connected to each other (left) and to
  the beam (right).  Contours are logarithmic, and the scales in the
  two figures are not the same.}
\label{fig:colorflow}
\end{center}
\end{figure}

Beyond kinematics, boosted perturbative decays can also differ from
QCD backgrounds in their color structure.  Consider a color singlet
such as a $H $ or $W $ boson decaying to a quark-antiquark pair.  The
daugher quark jets form a color dipole: they are color-connected to
each other, but not to the rest of the event. Meanwhile, the
backgrounds to these processes come from QCD dijets, which necessarily
have different color connections, as we show in
Fig.~\ref{fig:colorflow}, where the radiation patterns for a
color-singlet signal are plotted on the left and for a typical
background on the right, as computed in the eikonal (soft)
approximation. This observation has motivated work on variables which
can add color flow to the suite of features which can discriminate
signal from background.

%%%%%%%%%%%%%%%%%%%%%%%%%%%%%%%
%%%%%%%%%%%%%%%%%%%%%%%%%%%%%%%

\paragraph{Jet pull} \cite{Gallicchio:2010sw} defines for each jet a
transverse vector $\vec t_J$ characterizing the net directional
distribution of the soft radiation surrounding the jet core.  Defining
$\vec r_i$ as the (transverse) direction of particle $i$ from the jet
axis, the pull vector is
\beq
\vec{t}_J =\sum_{i\in J}\frac {p_{T,i}|r_i| \vec{r}_i}{p_{T,J}}.
\eeq
The direction of $\vec{t}_J$ relative to other jets in the event then
is sensitive to the color connection of the jet $J$.  Two jets which
are color-connected to each other will have pull vectors pointing
toward each other.  Jets which are color-connected to the beam will
have pull vectors pointing toward the beam.  Once two interesting
(sub)jets have been identified, the discriminating variable is then
$\cos\theta_t$, the angle between the pull vector and the line
connecting the two (sub)jet cores.
An initial experimental study of pull has been carried out at D0,
using  the $W$ in top events \cite{Abazov:2011vh}.

\paragraph{Dipolarity} \cite{Hook:2011cq} is a jet shape which is
designed to test for color dipole-like structure when the apparent
particle is boosted and the two (sub)jets of interest are
geometrically nearby. Since pull scales like $r_i^2$, it can be unduly
sensitive to the detailed assignment of particles between the two
(sub)jet cores to one or the other of the two (sub)jets.  Dipolarity
therefore uses as the relevant distance measure $R_i$, the transverse
distance of particle $i $ to the line segment connecting the (sub)jet
cores,
\beq
D_J = \frac{1}{p_{T,J} R_{12}^2} \sum_{i\in J} p_{T,i} R_i^2 .
\eeq
Note that dipolarity requires input (sub)jet axes.  The major
application studied to date has been in boosted top tagging, where
dipolarity can improve the identification of the boosted daughter $W$.

\paragraph{Keeping the right soft radiation.}
We have emphasized the need for jet grooming tools in the busy, high
luminosity environment of the LHC. However, that grooming will groom
away most if not all of the information about color flow.  To use the
information contained in an event's color flow, it is necessary to
retain at least some of the soft radiation. Exactly which soft
radiation is included, and at which stage in the analysis, is a
question which has to be addressed case-by-case.  As an example, we
will discuss how the dipolarity shape can be incorporated into a
boosted top tagger \cite{Hook:2011cq}.  Top tagging will be discussed
at length in the next section; for the moment, it suffices to think of
a top tagger as an algorithmic black box which acts on a fat jet to
return candidate $b$, $j_1$, and $j_2$ subjets, and discards some
radiation in the process.

The returned subjet axes define the characteristic opening scale,
$R_{12}$, and provide the input axes for the dipolarity jet shape.  As
the top-tagger has discarded some of the radiation associated with the
top quark in identifying the candidate subjets, to evaluate dipolarity
we will need to go back to the original fat jet and include a larger
subset of particles.  Clearly, the radiation we'd like to include when
evaluating the dipolarity of the candidate $W$ daughters is only that
associated with the two light quark jets; including radiation
originating from the $b$ would just skew the results. Let us consider
only moderately boosted tops, such that the $b$ jet is not overlapping
with the other two.  From the angular-ordered property of QCD showers,
we know that in top events, all radiation associated with either light
quark must be at angular separations less than the opening angle of
the dipole, $\Delta R<R_{12}$. Thus, all radiation from the $W$ is
contained in cones of radius $R_{12}$ around each light quark jet.
The authors of Ref.~\cite{Hook:2011cq} find that keeping all radiation
within these two cones is casting too wide a net, however, and a
smaller cone size of $R_{12}/\sqrt{2}$ is a better tradeoff between
keeping all the radiation from the $W$ and avoiding pollution from
pileup, underlying event, and splash-in from the nearby $b$.

Color flow variables capture a genuine physical difference between
signal and background.  They have been shown, in theoretical work, to
make a sizeable impact in signal significance \cite{Falkowski:2010hi,
  Cui:2010km, Hook:2011cq, Curtin:2012rm}, and show great promise as
tools to expand our understanding of SM and BSM physics.  It is
important to bear in mind, however, that these ``proof of principle''
analyses have all been performed using shower Monte Carlos, which
capture only leading approximations to the full QCD dynamics.  Just as
the jet shapes discussed in section~\ref{sec:raddist} above have been
and are still important tools for assessing the validity of the
approximations made in the Monte Carlo generators, measuring and
calibrating color flow variables in data is critical to understand the
validity of the shower models and the performance of any color flow
variable.  This experimental program is, as of yet, in its infancy.
In the meantime, theoretical studies should bear this uncertainty in
mind. To estimate the uncertainties, it is useful (as it is for any
novel substructure variable) to check results using more than one
shower model.

%%%%%%%%%%%%%%%%%%%%%%%%%%%%%%%%%%%%%%%%%%%%%%%%%%%%%%%%%%%%%%%%%%%%%%%%%%%%%
\section{Lecture III: Top tagging and searches for physics BSM}
%%%%%%%%%%%%%%%%%%%%%%%%%%%%%%%%%%%%%%%%%%%%%%%%%%%%%%%%%%%%%%%%%%%%%%%%%%%%%

In this section we will assemble the tools and techniques developed in
the previous two sections and apply them to searches for physics
beyond the standard model. By far the most universally motivated
application of jet substructure techniques to BSM physics is in the
hunt for TeV-scale new states which decay to electroweak-scale SM
particles.  The best reason for new physics to live anywhere near the
weak scale is that it is partially responsible for the generation of
the electroweak scale.  New physics that is related to EWSB will
naturally couple most strongly to those particles in the SM which feel
EWSB most strongly, in particular the top quark and the EW bosons
($H$, $W$, and $Z$), and thus will decay preferentially to these heavy
particles rather than to the light quarks and leptons which yield
simpler final states.  Moreover, we have compelling reasons to believe
new physics will naturally decay to {\it boosted} SM particles.
Even before the LHC turned on, the lack of deviations from SM
predictions for flavor or precision electroweak observables already
hinted that the likely scale for new physics was not $v_{EW}$ as
naturalness might have suggested, but rather $\Lambda \gtrsim $ few
TeV.  Evidence for this ``little hierarchy'' problem has of course
only gotten stronger as the LHC has directly explored physics at TeV
scales.  Thus many models which address the stabilization of the EW
scale will naturally give rise to final states rich in boosted tops,
Higgses, $W$'s and $Z$'s.

In this section we will provide an introduction to top tagging at the
LHC, followed by a few brief concluding comments on searching for more
general BSM physics with jets.

%%%%%%%%%%%%%%%%%%%%%%%%%%%%
\subsection{Top Tagging}
%%%%%%%%%%%%%%%%%%%%%%%%%%%%

As we established in Section 1, top pair production at the LHC covers
a broad range of kinematic regimes interpolating between threshold
($\sqrt{s}=2 m_t$), where tops are well described as a six-object
final state, up to TeV-scale energies, where the tops are highly
collimated and are best described as a two-object final state.  Top
reconstruction must thus be able to flexibly cover a wide range of
kinematic scenarios.  In the interest of time, we will restrict our
attention here to top taggers which target the hadronic decay of the
top quark, although the semi-leptonic decay mode also requires
interesting techniques for identification and reconstruction
\cite{Rehermann:2010vq, ATLAStop}.

As for jet algorithms, the ``best'' top tagger depends on the question
being asked.  In particular, different strategies are required at high
$p_T$ ($\gg m_t$) versus moderate $p_T$ ($\gtrsim m_t$).  Another
question is: what signal efficiency is necessary?  Every tagging
technique trades off signal efficiency against background mistag rate.
Depending on the search in question, the composition of the
backgrounds will change, and therefore the necessary mistag rate will
shift as well.  For example, consider a top pair event with at least
one boosted hadronic top.  If the other top is also hadronic, then QCD
dijets are by far the dominant background, and small QCD mistag rates
are required.  But if the other top is leptonic, then $W+$ jets
becomes an important background, and if the top is produced in
association with some new physics objects, such as $\met$, then the
backgrounds may be substantially smaller, and mistag rates may be
entirely unimportant.

The aim of this section is to provide an introduction to top tagging
by discussing a representative variety of top taggers.  Specifically,
we will consider the top taggers currently used by both LHC
experiments, which work best in the highly boosted regime; the ``HEP
top tagger'', which targets moderate $p_T$; and top tagging with
$N$-subjettiness.

\subsubsection{CMS top tagger}

The hadronic top tagger used by CMS \cite{CMS:2009lxa} is largely
based on the ``Hopkins'' top tagger \cite{Kaplan:2008ie}.  It builds
on the techniques of the boosted Higgs ``splitting/filtering'' or
``mass drop'' analysis, which we discussed in Section 1.  Thus, we
again begin by clustering the event using the C-A algorithm, on large
angular scales, capturing all of the top decay products in a single
fat jet, which we will then unwind until we find interesting
substructure.  Compared to the Higgs analysis, there are two important
differences.  First, we are looking for at least three hard subjets,
instead of two.  Second, we take the fat jet radius to be noticeably
smaller than we did for the Higgs case: $R = 0.8$.  Using our rule of
thumb, $R\sim 2 p_T/m$, this means we are targeting tops with
$p_T\gtrsim 500$ GeV: appropriate for production from a TeV-scale
resonance.  Contrast this with the boosted Higgs, which was targeting
the high-$p_T$ tails of SM associated production, where requiring
large $p_T$ imposed a significant price in signal acceptance.

Iteratively declustering the fat jet, we encounter splittings $P\to
ij$.  Our criterion for an interesting splitting is simply that both
daughter subjets must carry a sufficiently large fraction of the {\it total}
fat jet momentum,
\beq
p_{T,j}> \delta_P\, p_{T,J}
\eeq
for some parameter $\delta_P$.  If a splitting fails to meet this
criterion, discard the softer of $i,\, j$, and continue to unwind the
harder.  The splitting is rejected if it is too collinear, $|\Delta
\eta_{ij}|+|\Delta\phi_{ij} | > \delta_R$, for another parameter
$\delta_R$.  This procedure stops when either both $i$, $j$ are softer
than $\delta_P\, p_{T,J}$, or only one particle is left.

If an interesting hard, non-collinear splitting $P\to j_1 j_2$ is
found, then the next step is to successively unwind both $j_1$ and
$j_2$ according to the same algorithm, in search of further
interesting splittings.  This procedure returns a set of 2, 3, or 4
subjets.  Fat jets returning only 2 subjets don't have enough
substructure to be good top candidates, and are rejected.  Jets which
return 3 or 4 subjets do show enough substructure to be interesting,
and the next step is to test whether or not they also have top-like
kinematics.

As for the Higgs, the single most important discriminator is the jet
mass.  CMS requires that the jet mass, as computed from the sum of the
returned subjets, lie within a top mass window, $m_t-75\gev < m_J <
m_t+75$ GeV.  

The onshell decay of the $W$ inside the jet will also help us separate
signal from background, but rather than trying to explicitly identify
a pair of subjets which reconstruct a $W$---a procedure highly
vulnerable to the misassignment of particles in overlapping jets---we
will exploit the presence of the $W$ mass scale in a less direct way.

%%%%%%%%%%%%%%%%%%%%%%%%%%%%%%%
\begin{figure}
\begin{center}
\includegraphics[width=3.25in]{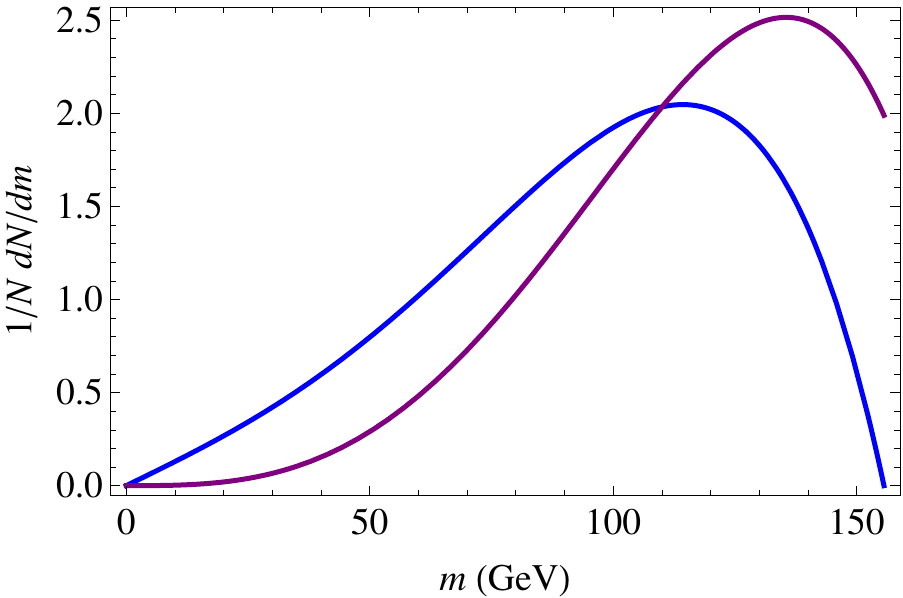}
\caption{Leading order distributions of $m_{b\bar d}$ (blue) and $m_{b
    u}$ (purple) in unpolarized top decay.}
\label{fig:mblmbnu}
\end{center}
\end{figure}
%%%%%%%%%%%%%%%%%%%%%%%%%%%%%%%

The pairwise invariant masses of all possible combinations of the
three daughter quarks are all governed by the mass scales in the top
matrix element, $m_W$ and $m_t$.  The distribution of the invariant
mass of the $b$ and the $\bar d$-type quark (equivalent to the charged
lepton in leptonic top decay) is shown in Fig.~\ref{fig:mblmbnu}. The
most likely value of $m_{b\bar d}$ is approximately 115 GeV.  The
invariant mass of the $b$ and the $u$-type quark (equivalent to the
neutrino) is peaked at even larger values. By contrast, subjet masses
from QCD background processes are hierarchically smaller than the
total parent jet mass.  Thus instead of trying to reconstruct the $W$,
we simply require that the {\it minimum} of the invariant masses
formed from pairs of the three hardest subjets be sufficiently large
to reject backgrounds,
\beq
\min (m_{12}, m_{13}, m_{23}) > 50\gev.
\eeq
These cuts on masses, together with the substructure requirement,
constitute the tagger. Note that no $b$-tagging information is
used. Tagging $b$-jets is very difficult in this environment for two
reasons. First, the $b$ is embedded in a highly collimated top, so
disentangling the tracks that are associated with the $b$ from the
other tracks in the jet is challenging. Second, the $b$ itself is at
very high $p_T$, so the opening angles of its daughter products are
small, and it is difficult to get sufficient resolution from the
reconstructed tracks to reconstruct the displaced vertex.  Note also
that the tagger doesn't require jet grooming.  This is partly because
the iterative decomposition procedure is performing some of that
function in its own right, as it discards soft wide-angle radiation in
the process of finding hard subjets (compare pruning).  The smaller
geometric size of the fat jets also means that pollution is not as
large an effect.

\subsubsection {ATLAS top tagger}

We turn next to the ATLAS top tagger.  Like CMS' tagger, it is
optimized for high $p_T$, and like CMS' tagger, it is based on
iterative declustering of a sequential algorithm. However, the ATLAS
tagger draws on a very different set of ideas, largely based on work
by Ref. \cite{Brooijmans:1077731} and the ``Y-splitter'' of
Ref. \cite{Butterworth:2002tt}.

The ATLAS top tagger begins by clustering events using the anti-$k_T$
algorithm with $R = 1.0 $.  (The slightly larger jet radius means that
this tagger works best at slightly lower $p_T$ than does the CMS
tagger.)  Since the anti-$k_T$ algorithm knows nothing about the
singularity structure of QCD, its use is simply to identify a nicely
regular initial set of particles.  The next step is to take this set
of particles and recluster them using the $k_T$ algorithm.

Recall that the $k_T$ algorithm preferentially clusters soft
splittings.  This means that the hardest splittings in the jet are the
very last ones.  Thus, there is no need to do any preliminary
unwinding, and the existence of hard substructure is directly
reflected in the hardness of the scales given by the $k_T$ metric
evaluated on the last few splittings in the jet:
\beq
d_{ij} = \min (p_{T,i}^2,p_{T, j}^2) \Delta R_{ij}^2.
\eeq
Large splitting scales mean the emissions are both hard and at wide
angles.  The ATLAS tagger uses as inputs the splitting scales of the last
three recombinations, $d_{12}$, $d_{23}$, and $d_{34}$.  The first two
splittings correspond (usually) to the identification of the three
daughter partons, and the third to possible FSR from one of the
partons.  Since for tops the splitting $d_{34}$ is the first which
comes from the QCD shower, its scale can still be relatively large; on
the other hand, for background QCD jets, the hierarchical nature of
the shower means that generally $d_{34}\ll d_{23} \ll d_{12}$.  Thus
cuts on $d_{34}$ maintain some discriminating power.

However, instead of cutting directly on the massive splittings
$d_{ij}$, it is advantageous to change variables to a set which are
less correlated with the jet and subjet invariant masses
\cite{Thaler:2008ju}.  We define the energy sharing variables
\beq
z_{ij}=\frac{d_{ij}}{d_{ij}+m_{ij}^2} \approx \frac{E_j}{E_i + E_j}
\eeq
where in the last step we have taken the collinear limit (and
$p_{T,i}>p_{T,j}$).  Notice that by performing this change of
variables we have removed sensitivity to the collinear singularity, so
that $z_{ij}$ is only capturing information about the soft
singularity.  Meanwhile, jet invariant masses still retain information
about the relative angles between the jets, so the correlation between
the variables has been reduced.

The final set of variables that make up the ATLAS top tagger is then:
\begin{itemize}

\item The total jet mass, $m_J$.  The tagger requires $m_J > 140$ GeV,
  and no upper bound: no grooming procedure is used, so the mass
  spectrum is distorted upwards.

\item The variable $Q_W$, defined as the minimum pair invariant mass
  of the three subjets identified at the splitting scale $d_{23}$.
  This is the equivalent to cutting on the minimum pair invariant mass
  in the CMS tagger; only the method of finding the subjets is
  different.  We require $Q_W>50$ GeV.

\item All three energy sharing variables, $z_{12}$, $z_{13}$, and
  $z_{23}$, which are subject to numerical cuts.

\end{itemize}

\subsubsection{HEP top tagger}

We turn now to the Heidelberg-Eugene-Paris top tagger, which functions
on tops with $p_T \gtrsim 200$ GeV \cite{Plehn:2009rk, Plehn:2010st}.
In some sense this algorithm is more of an event reconstruction
strategy than a top tagger.  The algorithm begins by clustering the
event using C-A on the extremely large angular scale $R=1.5$, and
requiring the fat jets thus formed to have $p_T> 200$ GeV.  The $p_T$
cut of 200 GeV puts us in the regime where the top is sufficiently
boosted that its decay products will frequently lie in a single
hemisphere.  Looking at extremely fat jets is effectively identifying
hemispheres in an event while avoiding the need to set any fixed
angular scales for resolution within those hemispheres.  This is an
effective strategy for tops in this intermediate kinematic regime,
where events will straddle any fixed angular scale; by unwinding C-A
hemispheres, we allow the angular scales to be flexibly identified
event by event.

The next step is to unwind the fat jet looking for interesting hard
structure.  This is done by employing a (loose) mass-drop criterion.
For a splitting $P\to ij$, with $m_j<m_i$, the splitting is deemed
sufficiently interesting if
\beq
\label{eq:HEPsplit}
m_j > 0.2\, m_P .
\eeq
If the splitting passes this criterion, retain both $i$ and $j$ in the
list of jets to unwind; otherwise, discard $j$ and keep unwinding $i$
until $m_i < 30$ GeV, at which point the unwinding stops.  This
unwinding procedure is performed on all subjets identified as
interesting via Eq.~\ref{eq:HEPsplit}.  The output of this step is a
list of subjets $\{j_i\}$ resulting from this iterative
declustering; if there are at least 3 such subjets, then we have found
enough substructure to continue.

At the next stage, we {\it filter} the substructures to shrink the
geometric area associated with the top daughters and thereby reduce
sensitivity to pileup, etc.  Unlike in the Higgs case, where the mass
drop criterion identified a unique angular scale $R_{b\bar b}$
associated with the sole hard splitting, we have a more complicated
set of jets with more than one interesting splitting, and it is not
immediately obvious which angular scale should be used to filter the
event.  The HEP top tagger determines the filter radius $R_{filt}$ by
brute force, as follows.  For each possible set of three subjets that
can be drawn from the $\{j_i\}$, filter them by resolving the
constituents of those subjets with radius $R_{filt}= \min(0.3, \Delta
R_{ij})$, and retain up to five subjets.  Let $m_{filt}$ be the invariant
mass of these up-to-five filtered subjets, and select the set with
$m_J$ closest to $m_t$ as the top candidate.  These up-to-five filtered
subjets are then (yet again) reclustered into three subjets, which are the
candidates for the partonic top daughters.  

The next step is to test whether or not the reconstructed top
daughters have top-like kinematics.  Again, we will exploit the
presence of both the top and $W$ mass scales.  We have already used
$m_t$ to identify the best set of subjets.  Unlike in the previous
taggers, we will now demand evidence of the on-shell $W$ in a more
complex way.  Label the three subjets returned by the previous step as
$\{ j_1, j_2, j_3\}$ in descending order of $p_T$. Of the three
invariant masses $m_{12}$, $m_{13}$, and $m_{23}$, only two are
independent. This means that the top kinematics is characterized by a
specific distribution in the two-dimensional space determined by the
pair invariant masses. Top jets are focused into a thin triangular annulus
in this space, as two subjets reconstruct an on-shell $W$ (the
annulus is triangular since any of the $m_{ij}$ may correspond to the
$W$).  Background, by contrast, is concentrated in regions of small
pairwise invariant masses.  The kinematic cuts imposed in the HEP top
tagger pick out this top-like triangular annulus by asking that events
lie on one of the three branches of the annulus.

%%%%%%%%%%%%%%%%%%%%%%%%%%%%%%%
\begin{figure}
\begin{center}
\includegraphics[width=4.25in]{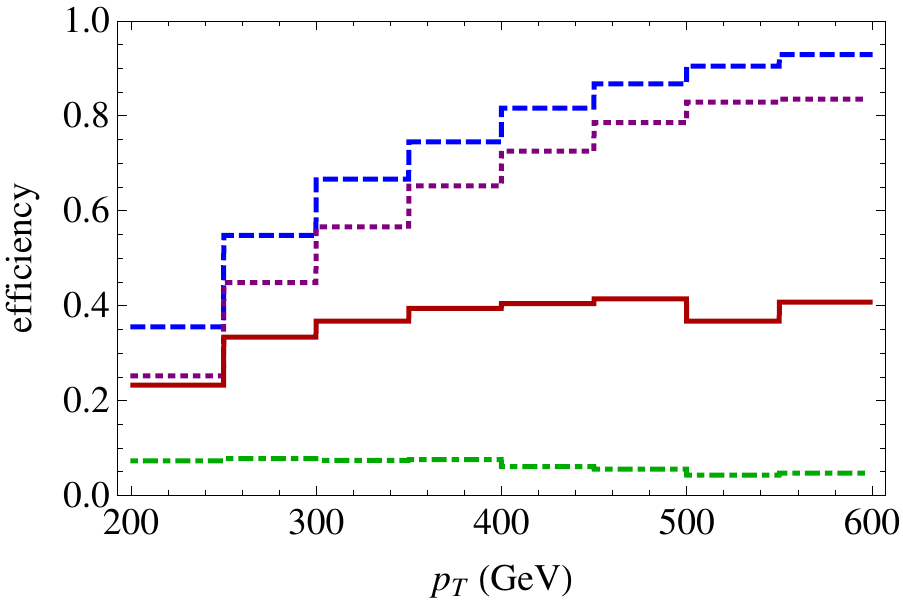}
\caption{HEP top tagger efficiencies on top quarks for: all top decay products
  within $\Delta R = 1.5$ of each other (blue, dashed); all top decay
  products clustered into $R= 1.5$ C-A fat jets (purple,
  dotted); tagged by the HEP top tagger (red, solid); tagged, but with
  reconstructed subjects not matching original partons (green,
  dot-dashed).  Data from Ref.~\cite{Plehn:2010st}.}
\label{fig:heptop}
\end{center}
\end{figure}
%%%%%%%%%%%%%%%%%%%%%%%%%%%%%%%
To understand how well this procedure covers the interpolating
kinematic region, we plot efficiencies for tops to pass through these
steps in Fig.~\ref{fig:heptop}.  As is evident from the blue (dashed)
curve, simply demanding that all decay products of the top lie in a
single hemisphere imposes a non-negligible acceptance price for tops
at the low end of the $p_T$ range, which drops quickly as the tops
become more energetic.  Further demanding that the top daughters all
be clustered into the same fat jet results in an additional mild
efficiency loss, seen in the purple (dotted) curve.  The purple curve
is the fraction of tops giving rise to taggable jets (neglecting the
possibility of mistagged signal).  The red line denotes the final
efficiency of the full HEP top tagger, after the filtering and
kinematic cuts.  At low $p_T$, the fraction of taggable jets which are
in fact tagged is near unity, but as the tops become more collimated,
the probability of a taggable jet passing the kinematic cuts falls
off, in large part because collimation and jet-particle misassignment
make the $W$ mass reconstruction less precise.  At the upper end of
the $p_T$ range shown in the figure, the high-$p_T$ top taggers are
useful, and would take over. Let us also comment that there is a
possibility for tops to pass the top tagger by accident, when the
algorithm picks up the wrong set of jets; this is shown in the green
(dot-dashed) curve.

\subsubsection{$N$-subjettiness}

As we saw in section~\ref{sec:shapesub}, $N$-subjettiness offers an
entirely complementary test of the existence of hard substructure. A
simple and highly effective top tagger can be constructed using as
input variables just the jet mass and the ratio $\tau_3/\tau_2$.
Further refinement is possible with a multivariate analysis which
uses in addition $\tau_2/\tau_1$ as well as $\tau_1,\, \tau_2,$ and
$\tau_3$ individually \cite{Thaler:2011gf}.  

From our experience with the previous taggers, we can guess that even
further improvement would be possible if some information about the
$W$ were also incorporated; since the $N$-subjettiness jet shape also
provides a method of determining subjet axes, it naturally suggests
methods for defining three subjets and computing the analog of $Q_W$.
To the best of the author's knowledge, no such study has been publicly
performed.

\subsubsection{Top tagging performance}

\begin{figure}
\begin{center}
\includegraphics[width=4in]{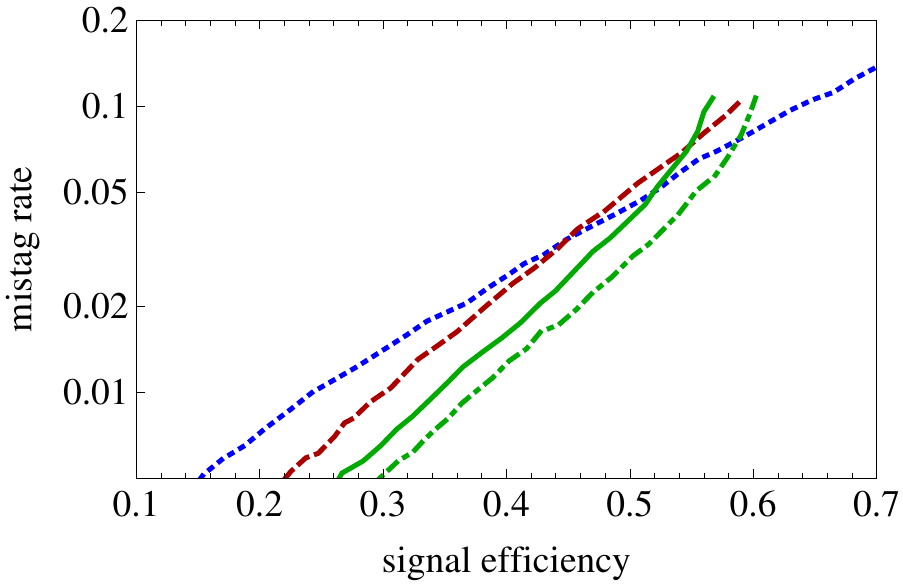}
\caption{ Top tagging performance curves, for tops with $200\gev < p_T
  < 800 \gev$ in the BOOST 2010 reference samples for: ATLAS (blue,
  dotted); CMS (red, dashed); and $N$-subjettiness in both the simple
  (green, solid) and multivariate (green, dot-dashed) versions.  Data
  from Refs.~\cite{BOOST2010,Thaler:2011gf}.}
\label{fig:eff1}
\end{center}
\end{figure}

Let us now consider the performance of the top taggers which we have
discussed.  This task is made easier by the work performed in the
BOOST 2010 \cite{BOOST2010} and BOOST 2011 \cite{BOOST2011} workshops,
which compared the performance of different top taggers on the same
reference sets of event samples.  These event samples are publicly
available online, so should you develop your own brilliant ideas about
top tagging, you can cross-check the performance of your novel
technique with the techniques already in the literature.  In
Fig.~\ref{fig:eff1} we show performance curves for the high-$p_T$ top
taggers which we discussed above. 
%%%%%%%%%%%%%%%%%%%%%%%%%%%%%%%
%%%%%%%%%%%%%%%%%%%%%%%%%%%%%%%
Overall, these high-$p_T$ top taggers have efficiencies on the order
of $\epsilon \sim 50\%$, at a (QCD) background mistag rate of
$\epsilon_{fake}\sim 5\%$.  (For comparison, LHC $b$-tagging
algorithms achieve $\epsilon \sim 70\%$, with a fake rate
$\epsilon_{fake}\sim 1\%$.)  It is evident from the performance curves
that the ATLAS tagger outperforms the CMS tagger when high signal
efficiency is required, while CMS does better at lower signal
efficiency.  Even the simple two-variable $N$-subjettiness tagger
outperforms both CMS and ATLAS taggers by a notable margin, except at
high signal efficiency, while adding the additional multivariate
discrimination to the $N$-subjettiness tagger provides a significant
improvement.  Further updates in the BOOST 2011 workshop show that (1)
being more precise about modeling QCD radiation at wide angles and (2)
including the effects of finite detector resolution reduce typical
efficiencies to $\epsilon \sim 40\%$, at a (QCD) background mistag
rate in the range $\epsilon_{fake}\sim 2-8\%$ depending on the tagger
\cite{BOOST2011}.  Incorporating finite detector resolution also tends
to reduce (but not erase!) the relative advantage of $N$-subjettiness
over the sequential decomposition-based taggers.

%%%%%%%%%%%%%%%%
\subsection{BSM searches with jet substructure}
%%%%%%%%%%%%%%%%

New physics produces jets with substructure when the kinematics
are governed by a nontrivial hierarchy of scales.  For the top
examples we've been discussing, this hierarchy arises from the
separation between the scale characterizing new physics and the
electroweak scale:
\beq
\Lambda_{NP} \gg \Lambda_{EW} \gg \Lambda_{QCD}.
\eeq
The little hierarchy problem results in a very strong motivation for
developing tagging techniques for boosted SM objects. Besides the top
tagging discussed in the previous subsection, much effort has also
gone into tagging boosted $W$, $Z$, and $H$ bosons arising from the
decay of new TeV-scale particles\cite{Kribs:2009yh, Katz:2010mr,
  Cui:2010km,Son:2012mb}. This is fortunate for theorists, as, once
these techniques are put into use at experiments in one context, the
barrier is much lower for their adaptation in other contexts where the
theoretical motivation may not be so universal.

What other kinds of BSM physics are amenable to substructure analyses?
To engender events with interesting substructure, some multi-tiered
hierarchy of scales is required.  We will enumerate an illustrative
but far from exhaustive set of examples.

Supersymmetry is one example of a new physics sector which naturally
can generate multiple scales. For example, if supersymmetry is broken
at very high scales, RG effects will drive the colored superpartners
much heavier than those superpartners with only EW charges.  Thus, at
the weak scale one could naturally expect $M_{\tilde g} \gg
M_{\chi^0}$.  In the presence of a large hierarchy between gluino and
neutralino, the decay products of the neutralino would be collimated.
Let us further suppose that the neutralino decays via the $R$-parity
violating $udd$ superpotential operator, so that $\chi^0\to q q q$.
Then gluino pair production would appear as a six-jet final state,
where two of the jets are actually boosted neutralinos, containing
interesting substructure \cite{Butterworth:2009qa}.  The very large
particle content of the MSSM can easily accommodate many possible
hierarchies, with different theoretical origins; see for instance
Ref.~\cite{Fan:2011jc} for another of the many possibilities.

Another way to generate a hierarchy in a BSM sector is if the new
physics sector contains a broken global symmetry, so that the scale
$\Lambda^{(1)}$ characterizing the lightest states is set by the
magnitude of the global symmetry breaking, rather than by the overall
scale of the new sector, $\Lambda^{(2)}\gg \Lambda^{(1)}$.  Thus
consider, for example, a composite rho $\rho_C$, decaying into two
pseudo-Nambu-Goldstone bosons $\pi_C$, which are stable within their
own sector, and therefore must subsequently decay into SM objects
\cite{Bai:2011mr}.

Hidden valley models also have this kind of multi-scale structure.
Here the hierarchy is between the mass of the mediator which connects
the visible and hidden sectors and the mass scale of the light states
in the hidden sector,
\beq
\Lambda_{med} \gg \Lambda_{NP} > \Lambda_{SM}.
\eeq
The mediating particle might be a SM particle, in particular the $H$
or $Z$, or a novel field such as a $Z'$ \cite{Baumgart:2009tn} or the
SM LSP \cite{Cheung:2009su}.  Exotic Higgs decays to light particles
also fall under this umbrella \cite{Chen:2010wk, Falkowski:2010hi,
  Kaplan:2011vf, Englert:2011iz, Lewis:2012pf, Draper:2012xt}.

More generally, thinking more broadly and flexibly about jets leads to
new approaches to combinatorics and event reconstruction
\cite{Hook:2012fd}, and provides novel methods to distinguish QCD
events from new physics.  As challenging high-multiplicity and
all-hadronic final states become a larger component of the LHC
program, flexible and creative jet techniques will be critical to our
ability to discover and interpret the physics.  Jet algorithms
themselves are still an evolving field! The anti-$k_T$ algorithm was
introduced only a few years ago.  As the nature of the questions that
we ask about jets evolves, so do the best jet algorithms to address
these questions.  There is still a lot of room for new ideas!

%%%%%%%%%%%%%%%%%%%%%%%%%%%%%%%%%%%%%%%%%%%%%%%%%%%%%%%%%%%%%%
\section{Further Reading}
%%%%%%%%%%%%%%%%%%%%%%%%%%%%%%%%%%%%%%%%%%%%%%%%%%%%%%%%%%%%%%

References which were invaluable in the preparation of these lectures
are the text {\it QCD and Collider Physics}, by Ellis, Stirling, and
Webber \cite{QCDnCP}, and the lecture notes ``Toward Jetography'' by
Salam \cite{jetography}.  The proceedings of the BOOST 2010 and 2011
workshops \cite{BOOST2010, BOOST2011} are valuable resources for those
looking for a quantitative survey of both theoretical and experimental
progress in jet physics at the Tevatron and the LHC.

\section*{Acknowledgments}  

It is a pleasure to thank M.~Schmaltz and the organizers for the
opportunity to be part of such an excellent program. I thank D.~Krohn
for introducing me to jet substructure, and for many useful
conversations during the course of our collaborations.  Thanks to
C.~Vermilion for providing Figure~\ref{fig:grooming}, and M.~Freytsis
and D.~Krohn for comments on the manuscript. Finally, thanks to my
collaborators, Y.~Bai, A.~Falkowski, A.~Thallapillal, and L-T.~Wang.
I am supported by DOE grant DE-FG02-92ER40704, NSF grant PHY-1067976,
and the LHC Theory Initiative under grant NSF-PHY-0969510.

\end{document}